\documentclass[conference]{IEEEtran}
\IEEEoverridecommandlockouts

\usepackage{cite}
\usepackage{amsmath,amssymb,amsfonts}
\usepackage{algorithmic}
\usepackage{graphicx}
\usepackage{textcomp}
\usepackage{xcolor}

\usepackage{acronym}
\usepackage{xcolor}

\usepackage{caption}
\usepackage{subfig}

\usepackage[hyphens]{url}
\usepackage{hyperref}
\hypersetup{breaklinks=true}
    
\usepackage[export]{adjustbox}
\usepackage{booktabs}
\usepackage{siunitx}

\usepackage{array}

\newcommand{\etal}{et al.}

\newcommand{\ie}{i.e.,}
\newcommand{\eg}{e.g.,}

\def\BibTeX{{\rm B\kern-.05em{\sc i\kern-.025em b}\kern-.08em
    T\kern-.1667em\lower.7ex\hbox{E}\kern-.125emX}}

\begin{document}

\title{
Practices for Engineering
\linebreak
Trustworthy Machine Learning Applications
}

\author{\IEEEauthorblockN{Alex Serban\IEEEauthorrefmark{1}\IEEEauthorrefmark{2}}
\IEEEauthorrefmark{1}ICIS, Radboud University,\\
\IEEEauthorrefmark{2}Software Improvement Group \\
a.serban@cs.ru.nl
\and
\IEEEauthorblockN{Koen van der Blom\IEEEauthorrefmark{3}, Holger Hoos\IEEEauthorrefmark{3} and Joost Visser\IEEEauthorrefmark{3}}
\IEEEauthorrefmark{3}LIACS, Leiden University,\\
The Netherlands
}

\maketitle

\begin{abstract}
Following the recent surge in adoption of \ac{ML},
the negative impact that improper use of \ac{ML} can have on users and society is now also widely recognised.
To address this issue, policy makers and other stakeholders, such as the European Commission or NIST,
have proposed high-level guidelines aiming to promote trustworthy \ac{ML}
(\ie~lawful, ethical and robust).
%
However, these guidelines do not specify actions to be taken by those involved in building \ac{ML} systems.
In this paper, we argue that guidelines related to the development of trustworthy \ac{ML} can be translated to operational practices, and should become part of the \ac{ML} development life cycle.
Towards this goal, we ran a multi-vocal literature review, and mined operational practices from white and grey literature.
Moreover, we launched a global survey to measure practice adoption and the effects of these practices.
In total, we identified 14 new practices, and used them to complement an existing catalogue of \ac{ML} engineering practices.
Initial analysis of the survey results reveals that so far, 
practice adoption for trustworthy \ac{ML} is relatively low.
In particular, practices related to assuring security of \ac{ML} components have very low adoption.
Other practices enjoy slightly larger adoption, such as providing explanations to users.
Our extended practice catalogue can be used by \ac{ML} development teams to bridge the gap between high-level guidelines and actual development of trustworthy \ac{ML} systems; it is open for review and contributions.
\end{abstract}


\acrodef{ML}[ML]{machine learning}
\acrodef{AI}[AI]{artificial intelligence}
\acrodef{RF}[RF]{random forest}
\acrodef{DL}[DL]{deep learning}
\acrodef{NN}[NN]{neural networks}
\acrodef{DNN}[DNN]{deep neural network}
\acrodef{SE}[SE]{software engineering}

\section{Introduction}
\label{sec:intro}

The recent increase in use of \acf{ML} to process personal data has resulted in more attention from policy makers and advisory bodies aiming to protect consumer interests.
A leading example is the independent high-level expert group on \ac{AI} set up by the European Commission~\cite{EUGuidelines}.
Well-intentioned but improper development of \ac{ML} components can cause unintentional harm~\cite{binns2018fairness}.
Published guidelines from policy makers  (\eg~\cite{EUGuidelines,national2019national,jobin2019global}) clearly reflect a desire for \ac{ML} components to be lawful, ethical and robust~\cite{EUGuidelines}.
However, these guidelines do not specify actions,
and usually come in the form of high-level checklists~\cite{EUChecklist} or broad recommendations~\cite{national2019national}.

In this paper, we aim to bridge the gap between guidelines from policy makers and operational 
practices for developers and their immediate collaborators.
The operational practices should become part of the \ac{ML} development life-cycle, together with established \ac{SE} practices.
We build on previous work by Serban~\etal~\cite{serban2020adoption}, which compiled a catalogue of \ac{SE} best practices for \ac{ML} applications.
While the catalogue covers a variety of \ac{SE} practices~--~\eg~testing of \ac{ML} components~--~it lacks practices in the areas of ethics and robustness.
These areas of interest are commonly grouped under the term \emph{trustworthy \ac{ML}}~\cite{EUGuidelines,brundage2020toward}.

We complement the catalogue of practices from~\cite{serban2020adoption} with 14 new practices for trustworthy \ac{ML}.
These practices are mined from literature through a multi-vocal review and tackle various topics, such as testing for bias, assuring security, and
having the application audited by third parties.
For all practices, we summarise related work into an actionable body of knowledge including the practice intent, motivation, applicability, description and references.
We also launched a global survey to measure adoption of the practices, and assess their effects.
In this paper, we describe the survey and present early results.
Moreover, we invite the community to contribute to the catalogue of practices, and more generally to a body of knowledge spanning trustworthy \ac{ML} and \ac{SE}.

We begin by discussing background information and related work (Section~\ref{sec:background}). 
Following that, we present the process of mining practices from literature, the resulting practices, and early results from the survey (Section~\ref{sec:se_ml}).
We close with a discussion and concluding remarks (Section~\ref{sec:discussion}).

\section{Background and Related Work}
\label{sec:background}

The literature on \ac{SE} for \ac{ML} consists of challenges faced by practitioners in the adoption of \ac{ML}~\cite{arpteg2018software}, practices~\cite{serban2020adoption}, guidelines~\cite{sculley2015hidden}, and design patterns~\cite{washizaki2019studying}.
These lines of work tackle issues related to the \ac{ML} development process, which includes rapid iterations by multi-disciplinary teams~\cite{arpteg2018software}, an experimental approach to software development~\cite{sculley2015hidden}, the need to tackle uncertainty of \ac{ML} components~\cite{serban2020towards}, and a strong emphasis on automation of operational aspects~\cite{serban2020adoption}.

However, none of these lines of work tackle issues related to the negative impact that improper use of \ac{ML} has on society.
For example, inappropriate testing for bias can have devastating effects on some social groups~\cite{binns2018fairness}.
Public documents from policy makers and advisory bodies~--~such as the independent high-level expert group on \ac{AI} set up by the European Commission~\cite{EUGuidelines}~--~acknowledge the potential negative impact of \ac{ML} on society, and propose guidelines to address these issues. 
However, the recommended guidelines are not immediately operational, and need substantial interpretation and refinement to specific actions in a development context.

In order to complement guidelines from policy makers, increase their operational applicability and relate them to established engineering practices, we mined practices for trustworthy \ac{ML} from the literature using academic (white) and non-academic (grey) literature.
The latter is known to benefit \ac{SE} research by providing valuable information from experience~\cite{garousi2019guidelines}, and its value in the context of \ac{SE} for \ac{ML} has been demonstrated before~\cite{serban2020adoption}.

\section{Practices for Trustworthy Machine Learning}
\label{sec:se_ml}

\textbf{Practice mining from literature.}
To extract the practices, we followed a similar process to~\cite{serban2020adoption}, \ie{} we ran a multi-vocal literature review on the topic of trustworthy \ac{ML}.
As sources of information we used Google (for grey literature search) and Google Scholar (for both white and grey literature search).
We composed queries with synonyms or disambiguation of \emph{trustworthy \ac{ML}}.
The queries were formed from two elements; the first suggesting the field of research, and the second suggesting sub-fields of trustworthy \ac{ML}.
For the first element, we used three possible variations~--~machine learning, deep learning, and \ac{AI}.
For the second element, we used multiple variations, inspired by~\cite{EUGuidelines}~--~robustness, privacy, fairness, bias, interpretable, transparent, ethical, and auditable.
As an example, the first query was \emph{machine learning robustness}.

Exclusion criteria were formulated to avoid duplicates and articles published before 2016, which are generally subsumed by later articles~\cite{serban2020adoption}.
We included all articles which proposed challenges, requirements, practices or future directions on the topics summarised in the second element of the query.
Moreover, we included all relevant documents from policy makers, as well as papers describing their content.

When interpreting the query results, we defined practices by identifying and merging common themes between challenges and solutions.
The challenges provided the intent and motivation, and the solutions provided the description of the practice. 
This procedure is known as thematic analysis~\cite{braun2006using}, and it is commonly used in qualitative \ac{SE} studies.
When compiling the practices, we emphasised their applicability and their match to different stages of the \ac{ML} development process.

\textbf{Resulting practices for trustworthy \ac{ML}.}
In total, we identified 13 relevant articles, from which we compiled 14 new practices for trustworthy \ac{ML}.
Moreover, we included two additional articles, that better position the practices, or bridge \ac{ML} and \ac{SE}~\cite{ruhe2002software,branke2008multiobjective}.
To classify the practices, we used the taxonomy from~\cite{serban2020adoption}, who provided evidence that a common taxonomy for the \ac{ML} development process does not exist, and reconstructed a general taxonomy that is compatible with previous work.
Since our work complements their catalogue of practices, it is natural to use the same taxonomy.

\begin{table}[t]
    \caption{Practices for trustworthy \ac{ML}, along with the requirements (Req.) for trustworthy AI from~\cite{EUGuidelines}, that they address. The requirements are (R1) human agency and oversight, (R2) technical robustness and safety, (R3) privacy and data governance, (R4) transparency, (R5) diversity, non-discrimination and fairness, (R6) societal and environmental well being, and (R7) accountability.}
    \label{tbl:practices}
    \centering
    \begin{tabular}{p{0.17cm}p{4.2cm}p{0.4cm}p{0.3cm}p{1.8cm}} 
        \toprule
         
         \textbf{Nr.} & \textbf{Title} & \textbf{Class} & \textbf{Req.} & \textbf{References} \\ 
        
        \midrule
        T1 & Test for social bias in training data & Data & R5 & \cite{FairMH,liu2019implicit,kleinberg2016inherent,zemel2013learning} \\ 
        
        T2 & Prevent discriminatory data attributes from being used as model features & Data & R5 & \cite{FairMH} \\
        
        T3 & Use privacy preserving \ac{ML} techniques & Data & R3 &  \cite{brundage2020toward} \\ 
        
        T4 & Employ interpretable models whenever possible & Tr. & R1 &  \cite{brundage2020toward,EUGuidelines,molnar2020interpretable} \\ 
        
        %
        
        T5 & Assess and manage subgroup bias & Tr. & R5 &  \cite{hebert2018multicalibration,kearns2018preventing} \\ 
        
        T6 & Assure application security & Code & R2 &  \cite{brundage2020toward,national2019national} \\ 
        
        T7 & Provide audit trails & Dep. &  R7 &  \cite{EUGuidelines,brundage2020toward,raji2020closing} \\ 
        
        T8 & Decide trade-offs through an established team process & Tm. &  R1 & \cite{fandel1990group,branke2008multiobjective,ruhe2002software} \\ 
        
        T9 & Establish responsible \ac{AI} values & Gov. & R4 &  \cite{EUGuidelines,brundage2020toward} \\ 
        
        T10 & Perform risk assessments & Gov. & R2 &  \cite{EUGuidelines,brundage2020toward,raji2020closing}  \\

        T11 & Inform users on \ac{ML} usage & Gov. & R4 &  \cite{EUGuidelines,mitchell2019model} \\

        T12 & Explain results and decisions to users & Gov. & R4 &\cite{EUGuidelines}  \\

        T13 & Provide safe channels to raise concerns & Gov. & R4 & \cite{EUGuidelines,brundage2020toward} \\

        T14 & Have your application audited & Gov. & R7 & \cite{EUGuidelines,brundage2020toward} \\ 
        \bottomrule
        
    \end{tabular}
\end{table}

Our list of practices, together with their class and references, is shown in Table~\ref{tbl:practices}.
We further discuss the practices and their relationship to the requirements for trustworthy AI from~\cite{EUGuidelines}, which is widely regarded as mature and authoritative.
We plan to discuss the relationship of the practices with other high-level guidelines in future work.

The practices in the Data class describe topics such as testing for social bias, preventing discriminatory attributes from being used as inputs to \ac{ML} components, and using privacy-preserving techniques.
These practices address the requirements for \emph{privacy and data governance}, and for \emph{diversity, non-discrimination and fairness} from~\cite{EUGuidelines}.

Testing for bias is also present in the Training (Tr.) class, where it refers to assessing whether a trained model exhibits subgroup bias, which may arise from balancing groups in the training data.
Moreover, some techniques for managing subgroup bias involve changing the loss function of 
\ac{ML} components, or performing calibrations~\cite{hebert2018multicalibration}. 
These techniques are applied when training a model, and not during data preparation.
Another practice for training relates to using interpretable models whenever possible.
The Training practices address the requirements for \emph{human agency and oversight}, and for \emph{diversity, non-discrimination and fairness} from~\cite{EUGuidelines}.

In the Coding (Code) class, there is a single practice related to ensuring security.
This practice tackles \ac{ML} security topics, such as robustness, and more traditional security concerns, such as penetration testing.
The practice addresses the requirement for \emph{technical robustness and safety} from~\cite{EUGuidelines}.

In the Deployment (Dep.) class, there is a single practice related to logging and storing audit trails. 
This practice addresses the requirement for \emph{accountability} from~\cite{EUGuidelines}.

We also identified a single practice in the Team (Tm.) class, related to defining team processes for deciding on \ac{ML}-specific trade-offs. For example, processes for deciding on prioritisation of interpretable \ac{ML} methods over non-interpretable methods, on prioritisation of non-biased models with lower accuracy, or on whether to use \ac{ML} in the first place.
This practice addresses the requirement for \emph{human agency and oversight}  from~\cite{EUGuidelines}.

The largest class of practices~--~Governance (Gov.)~--~covers organisational practices which guide the development of \ac{ML} components.
Examples of such practices include the adoption of responsible \ac{AI} values, subscription to a code of conduct, risk assessments, and auditing of \ac{ML} components by third parties.
Moreover, the Governance class includes a series of practices based on enforceable laws, such as explaining results and decisions to users (\ie{} satisfying the right to an explanation).
These practices address the \emph{accountability} and \emph{transparency} requirements from~\cite{EUGuidelines}.

\begin{figure}[t]
    \fbox{
    \parbox{8.5cm}{%
    \textbf{Test for social bias in training data} 
    \hspace*{2cm} \textbf{\emph{Data}} \\  
    \\ 
    \textbf{Intent} \\ Identify instances of social bias in training data, [...]. \\ 
    \textbf{Motivation}  \\
    Bias in data is one of the main sources of unfairness [...]. \\
    \textbf{Applicability} \\
    Testing for social bias in training data should be done 
    [...]. \\ 
    \textbf{Description} \\ 
    In order to avoid social bias in ML algorithms,
    it is imperative to continuously check that the training data [...]. 
    \\
    \textbf{Adoption} \\ 
    Adoption rates, grouped by distinct demographic factors.
    
    \textbf{Related} \\ 
    Practices 2, 5, and 10 from Table~\ref{tbl:practices}. \\ 
    \textbf{References} \\  \cite{FairMH,liu2019implicit,kleinberg2016inherent,zemel2013learning} 
    }}
    \caption{Example of practice in online catalogue.}
    \label{fig:practice_example}
\end{figure}

To make the practices available to practitioners, we added them to an existing online catalogue\footnote{\url{https://se-ml.github.io/practices/}} (previously built by Serban~\etal~\cite{serban2020adoption}), consisting of detailed descriptions and concise statements of intent, motivation, applicability, related practices and references.
A brief (and abbreviated) example is illustrated in Figure~\ref{fig:practice_example}.
A curated reading list with these references, further relevant literature, as well as a selection of supporting tools is maintained online\footnote{\url{https://github.com/SE-ML/awesome-seml}}.
Both catalogue and reading list are open for community contributions.

\textbf{Practice adoption.}
In order to measure the adoption of the practices, we extended the survey from~\cite{serban2020adoption} with new questions for the practices in Table~\ref{tbl:practices}.
The extended survey allows joint assessment of practices for trustworthy \ac{ML} and more established \ac{ML} engineering practices from~\cite{serban2020adoption}.
The survey's questionnaire was designed following the recommendations 
of Kitchenham et al.~\cite{kitchenham2002principles2}.
It is a cross-sectional observational (\ie~participants were asked at the moment of taking the questionnaire to what extent they adopted the practices) concurrent control study (\ie~participants are assigned to specific groups, enabled by several preliminary questions).

Besides 14 new questions with the practices in Table~\ref{tbl:practices}, we added one question regarding the effect of adopting practices for trustworthy \ac{ML}.
The answers were standardised on a Likert scale, with four possible answers, designed to express degrees of adoption~--~\eg~``not at all'' or ``completely''~--~rather than degrees of agreement such as ``agree'' or ``strongly agree''.
When the answer scale did not match the full range of possible answers, we added specific answers that helped to avoid noisy results (\ie~variations of ``not applicable'').

\begin{figure}
    \centering
    \includegraphics[width=9cm,keepaspectratio]{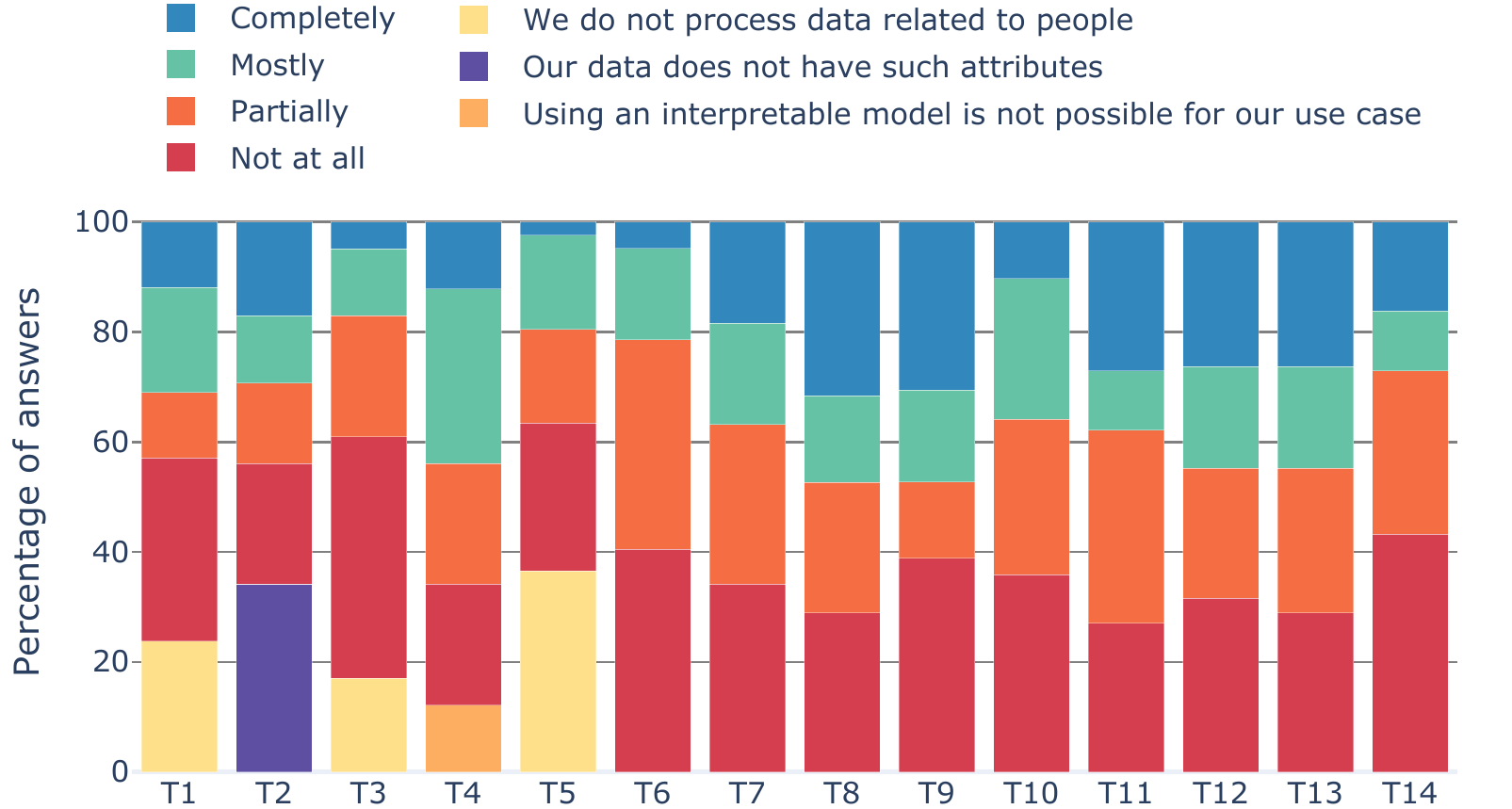}
    \caption{Adoption of \ac{SE} practices for trustworthy \ac{ML}.}
    \label{fig:practice_adoption}
\end{figure}

\textbf{Practice adoption results.}
In the time span of three months, we received 103 answers, from which we filtered out participants that were not part of a team using \ac{ML}.
Moreover, we filtered on the percentage of questions that were answered in the prerequisites (at least \SI{50}{\percent}) and in the practice adoption questions (at least \SI{50}{\percent}), resulting in \emph{42~complete responses}.

Figure~\ref{fig:practice_adoption} presents our preliminary results on practice adoption. 
The first 5 practices had specific answers to avoid noisy results, such as ``using interpretable models is not possible for our use case"~--~which accounts for use cases where black box models from deep learning are heavily used (\eg~computer vision).
Moreover, for questions related to social bias, we added an answer specifying that the data set does not contain personal data, and thus the practices are not applicable (``we do not process data related to people'').
Similarly, for using discriminatory attributes, the participants were allowed to specify that the data sets used do not contain such attributes, through the ``our data does not have such attributes" answer.

More than \SI{20}{\percent} of the participants reported that they do not use personal data in their projects, or data with discriminatory attributes (T1, T2, T5).
For practice T3~--~\ie~use privacy-preserving \ac{ML} techniques~--~a smaller percentage of participants chose the extra answer.
We hypothesise that participants found the ``Not at all" answer a better fit for this question.

Overall, the practice adoption for trustworthy \ac{ML} is rather low, based on the large percentage of ``Not at all" and ``Partially" answers.
The least adopted practice is T6, on assuring application security.
This result is cause for concern, given the interest in both software and \ac{ML} security from academia and industry.
We hypothesise that although a large body of academic literature regarding security and robustness of \ac{ML} exists, it is still limited in its applicability.
For example, all defences against adversarial examples~--~a known threat for \ac{ML} components~--~have been breached~\cite{serban2020adversarial}.
This practice is also linked to practice T10, on performing risk assessments, which also has low adoption.
Following that is practice T14, which involves having the application audited by third parties.

At the other end of the spectrum, the practices related to establishing team processes for deciding trade-offs (T8) and establishing responsible AI values (T9) have higher adoption.
Similarly, practice T12 on explaining results to users, practice T13 on providing safe channels to raise concerns, and practice T4 on employing interpretable models have also have slightly higher adoption.
We hypothesise that larger adoption of some practices (T4, T12, T13) is motivated by more mature legislation in the area (\eg~the right to an explanation).
Other practices are more established in the engineering community (T8), or have higher practical feasibility (T9). 

\section{Discussion and concluding remarks}
\label{sec:discussion}

We introduced a set of 14 new practices for trustworthy \ac{ML}, which complements the existing catalogue of \ac{ML} engineering practices from~\cite{serban2020adoption}.
We argue that requirements related to the development of trustworthy \ac{ML} (outlined by policy makers and regulatory bodies) can be translated into operational practices and should become part of the \ac{ML} development life cycle.
We also believe that the adoption of trustworthiness-specific and general ML engineering practices is interconnected; 
for instance, the practice of continuous integration~\cite{serban2020adoption} can make the practices for bias testing more effective. 
Using results from our extended survey, we plan to study the joint adoption of practices and their effects in more detail.

We emphasise that the practices introduced in this paper are meant to complement, not replace, the guidelines from policy makers and advisory bodies.
We also note that no practices that directly address requirement R6 from~\cite{EUGuidelines} could be identified through our literature review.
While practices (T1-T3, T9) address the societal aspects of R6,
the environmental well-being is only addressed indirectly by T9.
In the future, we plan to test the completeness of our catalogue through validation interviews with practitioners. 
Moreover, we welcome 
contributions from the community to further enrich the catalogue.

The analysis of the initial answers from our survey revealed low adoption of practices for trustworthy \ac{ML}.
In the future, we plan to further increase the number of respondents.
New responses will also enable fine-grained analyses of the data~--~\eg{} conditional analysis by demographics, correlation among practices, and prediction of effects from practice adoption. 
By repeating the survey at regular intervals, we will be able to monitor future adoption trends.

While in this paper we focused on the high-level requirements for trustworthiness from~\cite{EUGuidelines}, similar guidance documents exist, or are being developed. 
We believe it is important to perform a broad and deep study of such documents, to identify where additional operational practices need to be provided.
We also plan to conduct interviews with practitioners to identify new practices, that might not be described in literature, to further reduce the gap between policy and practice.

In this work we measured practice adoption through self-assessment by practitioners. A challenge for future work is to complement this with more objective, evidence based methods.

\bibliographystyle{ieeetr}
\bibliography{bibliography}

\begin{thebibliography}{10}

\bibitem{EUGuidelines}
{High-Level Expert Group on AI}, ``Ethics guidelines for trustworthy {AI}.''
  \url{https://ec.europa.eu/digital-single-market/en/news/ethics-guidelines-trustworthy-ai}.
\newblock [Online; accessed 14-01-2021].

\bibitem{binns2018fairness}
R.~Binns, ``Fairness in machine learning: Lessons from political philosophy,''
  in {\em Conference on Fairness, Accountability and Transparency}, vol.~81 of
  {\em PMLR}, PMLR, 23--24 Feb 2018.

\bibitem{national2019national}
{National Science and Technology Council (US). Select Committee on Artificial
  Intelligence}, ``The national artificial intelligence research and
  development strategic plan: 2019 update.''
  \url{https://www.nitrd.gov/news/National-AI-RD-Strategy-2019.aspx}.
\newblock [Online; accessed 14-01-2021].

\bibitem{jobin2019global}
A.~Jobin, M.~Ienca, and E.~Vayena, ``The global landscape of {AI} ethics
  guidelines,'' {\em Nature Machine Intelligence}, vol.~1, no.~9, 2019.

\bibitem{EUChecklist}
{High-Level Expert Group on AI}, ``Assessment list for trustworthy artificial
  intelligence for self-assessment.''
  \url{https://ec.europa.eu/digital-single-market/news/assessment-list-trustworthy-artificial-intelligence-altai-self-assessment}.
\newblock [Online; accessed 14-01-2021].

\bibitem{serban2020adoption}
A.~Serban, K.~van~der Blom, H.~Hoos, and J.~Visser, ``Adoption and effects of
  software engineering best practices in machine learning,'' in {\em
  International Symposium on Empirical Software Engineering and Measurement
  (ESEM)}, ACM, 2020.

\bibitem{brundage2020toward}
M.~Brundage {\em et~al.}, ``Toward trustworthy {AI} development: Mechanisms for
  supporting verifiable claims,'' {\em arXiv:2004.07213}, 2020.

\bibitem{arpteg2018software}
A.~Arpteg, B.~Brinne, L.~Crnkovic-Friis, and J.~Bosch, ``Software engineering
  challenges of deep learning,'' in {\em Euromicro Conference on Software
  Engineering and Advanced Applications (SEAA)}, IEEE, 2018.

\bibitem{sculley2015hidden}
D.~Sculley, G.~Holt, D.~Golovin, {\em et~al.}, ``Hidden technical debt in
  machine learning systems,'' in {\em NeurIPS}, 2015.

\bibitem{washizaki2019studying}
H.~Washizaki {\em et~al.}, ``Studying software engineering patterns for
  designing machine learning systems,'' in {\em IWESEP}, IEEE, 2019.

\bibitem{serban2020towards}
A.~Serban {\em et~al.}, ``Towards using probabilistic models to design software
  systems with inherent uncertainty,'' in {\em ECSA}, Springer, 2020.

\bibitem{garousi2019guidelines}
V.~Garousi, M.~Felderer, and M.~V. M{\"a}ntyl{\"a}, ``Guidelines for including
  grey literature and conducting multivocal literature reviews in software
  engineering,'' {\em Information and Software Technology}, vol.~106, 2019.

\bibitem{braun2006using}
V.~Braun and V.~Clarke, ``Using thematic analysis in psychology,'' {\em
  Qualitative research in psychology}, vol.~3, no.~2, 2006.

\bibitem{ruhe2002software}
G.~Ruhe, ``Software engineering decision support--a new paradigm for learning
  software organizations,'' in {\em International Workshop on Learning Software
  Organizations}, Springer, 2002.

\bibitem{branke2008multiobjective}
J.~Branke, K.~Deb, K.~Miettinen, and R.~Slowi{\'n}ski, {\em Multiobjective
  optimization: Interactive and evolutionary approaches}, vol.~5252 of {\em
  Lecture Notes in Computer Science}.
\newblock Springer, 2008.

\bibitem{FairMH}
M.~Hardt, ``Fairness.'' \url{https://www.youtube.com/watch?v=Igq_S_7IfOU},
  2020.
\newblock [Online; accessed 14-01-2021].

\bibitem{liu2019implicit}
L.~T. Liu, M.~Simchowitz, and M.~Hardt, ``The implicit fairness criterion of
  unconstrained learning,'' in {\em ICML}, PMLR, 2019.

\bibitem{kleinberg2016inherent}
J.~Kleinberg, S.~Mullainathan, and M.~Raghavan, ``Inherent trade-offs in the
  fair determination of risk scores,'' {\em arXiv:1609.05807}, 2016.

\bibitem{zemel2013learning}
R.~Zemel, Y.~Wu, K.~Swersky, T.~Pitassi, and C.~Dwork, ``Learning fair
  representations,'' in {\em ICML}, PMLR, 2013.

\bibitem{molnar2020interpretable}
C.~Molnar, {\em Interpretable Machine Learning}.
\newblock Lulu.com, 2020.

\bibitem{hebert2018multicalibration}
{\'U}.~H{\'e}bert-Johnson {\em et~al.}, ``Multicalibration: Calibration for the
  (computationally-identifiable) masses,'' in {\em ICML}, PMLR, 2018.

\bibitem{kearns2018preventing}
M.~Kearns {\em et~al.}, ``Preventing fairness gerrymandering: Auditing and
  learning for subgroup fairness,'' in {\em ICML}, PMLR, 2018.

\bibitem{raji2020closing}
I.~D. Raji {\em et~al.}, ``Closing the {AI} accountability gap: defining an
  end-to-end framework for internal algorithmic auditing,'' in {\em Conference
  on Fairness, Accountability, and Transparency}, PMLR, 2020.

\bibitem{fandel1990group}
G.~Fandel, ``Group decision making: Methodology and applications,'' in {\em
  Readings in Multiple Criteria Decision Aid}, Springer, 1990.

\bibitem{mitchell2019model}
M.~Mitchell {\em et~al.}, ``Model cards for model reporting,'' in {\em
  Conference on Fairness, Accountability, and Transparency}, PMLR, 2019.

\bibitem{kitchenham2002principles2}
B.~A. Kitchenham and S.~L. Pfleeger, ``Principles of survey research part 2:
  designing a survey,'' {\em ACM SIGSOFT Software Engineering Notes}, vol.~27,
  no.~1, 2002.

\bibitem{serban2020adversarial}
A.~Serban, E.~Poll, and J.~Visser, ``Adversarial examples on object
  recognition: a comprehensive survey,'' {\em ACM CSUR}, vol.~53, no.~3, 2020.

\end{thebibliography}
\end{document}